\documentclass[prl,aps,10pt,twocolumn,showkeys,nofootinbibs,showpacs]{revtex4-1}

\usepackage{amsmath,amssymb,amsfonts,amsthm}
\usepackage{amsbsy} 
\usepackage{epsfig}
\usepackage{latexsym}
\usepackage[utf8]{inputenc}
\input amssym.def
\input amssym.tex

\usepackage{hyperref}

\newcommand{\oarX}[1]{\href{http://arxiv.org/abs/#1}{{\ttfamily #1}}}
\newcommand{\arX}[1]{\href{http://arxiv.org/abs/#1}{{\ttfamily arXiv:#1}}}

\def\barr{\begin{array}}
\def\earr{\end{array}}
\def\half{\frac{1}{2}}
\def\ben{\begin{equation}}
\def\een{\end{equation}}
\def\bs{\begin{subequations}}
\def\es{\end{subequations}}
\def\bena{\begin{eqnarray}}
\def\eena{\end{eqnarray}}

\def\im{{\rm i}}

\def\be{\begin{equation}}
\def\ee{\end{equation}}

\def\bes{\begin{eqnarray}}
\def\ees{\end{eqnarray}}

\begin{document}

\title{Perfect Quantum Cosmological Bounce}

\author{Steffen Gielen}
\affiliation{Theoretical Physics, Blackett Laboratory, Imperial College London, London SW7 2AZ, United Kingdom}
\author{Neil Turok}
\affiliation{Perimeter Institute for Theoretical Physics, Waterloo, Ontario N2L 2Y5, Canada}

\date{\today}


\begin{abstract}
We study quantum cosmology with conformal matter comprising a perfect radiation fluid and a number of conformally coupled scalar fields. Focusing initially on the collective coordinates (minisuperspace) associated with homogeneous, isotropic backgrounds, we are able to perform the quantum gravity path integral exactly. The evolution describes a ``perfect bounce'', in which the Universe passes smoothly through the singularity. We extend the analysis to spatially flat, anisotropic universes, treated exactly, and to generic inhomogeneous, anisotropic perturbations treated at linear and nonlinear order. This picture provides a natural, unitary description of quantum mechanical evolution across a cosmological bounce. We provide evidence for a semiclassical description in which all fields pass ``around'' the cosmological singularity along complex classical paths. 
\end{abstract}

\pacs{98.80.Qc,~04.60.Kz,~98.80.Cq,~04.20.Dw}

\maketitle

Recent observations have revealed extraordinary simplicity in the large-scale structure of the Universe: a spatially flat geometry with nearly scale-invariant, Gaussian fluctuations.  As yet, there are no indications of tensor (gravitational wave) modes which would signal primordial inflation, nor complications such as curvature,  non-Gaussianity, or isocurvature modes. The simplicity of these findings seems at odds with expectations based on inflationary models predicting a ``multiverse'' with random and unpredictable behavior on large scales.  We are therefore encouraged to seek more economical explanations for the state of the Universe. One of the oldest and simplest ideas~\cite{Friedmann} is that the big bang was a bounce. Such a bounce is generally forbidden in classical general relativity, but might be allowed in quantum gravity~\cite{LoopBounce}. 

At the big bang singularity, the density and temperature of matter diverges. General arguments indicate that the only complete quantum field theories are those with a UV fixed point, i.e., which are conformally invariant at high energies~\cite{Shomer}. Therefore, it is of particular interest to study cosmologies with conformally invariant matter. A simple example is a spatially  flat, homogeneous, and isotropic universe filled with a perfect radiation fluid,  with line element
\ben
ds^2=a^2(\eta)(-d\eta^2+d\vec{x}^2), 
\label{eq0}
\een
with $a(\eta)\propto \eta$, where $\eta$ is the conformal time. This is a good metric for all $\eta\neq 0$, and furthermore possesses a unique analytic continuation around the singularity in the complex $\eta$-plane. As we shall see, generic perturbations about this metric share these nice properties; hence, we term this cosmology a ``perfect bounce". Indeed, in a pioneering paper~\cite{dewitt} [see discussion following Eq.~(5.19)], DeWitt anticipated this idea, stating: ``One might hope that an analytic continuation could be performed around [the singularity] but whether this would have any physical meaning is unclear." We shall perform just such a continuation and interpret it. 

We study the quantum dynamics of a universe with conformal matter consisting of perfect radiation and conformally coupled scalar fields. We first analyze homogeneous, isotropic backgrounds, showing that the semiclassical approximation to the path integral is exact. We then generalize to {\it anisotropic} spatially flat metrics, computing the Feynman propagator and clarifying its analytic properties. Finally, we tackle generic {\it inhomogeneous} cosmologies, order by order in a perturbative expansion. Although perturbation theory fails as we approach $\eta=0$, we can maintain its validity by deforming the $\eta$-contour into the complex $\eta$-plane and bypassing the singularity. This continuation respects all the symmetries of general relativity and yields an unambiguous result. For conformal matter, we show there is no quantum creation of scalar density perturbations or gravitational waves across the bounce, indicating a well-defined vacuum. For theories with nontrivial running and/or soft breaking terms, one would find finite particle production. 

{\em Weyl-invariant formulation.}---We consider cosmology with perfect radiation and $M$ conformally coupled scalars $\vec{\chi}=(\chi^1,\dots,\chi^M)$. It is convenient to ``lift'' Einstein gravity to a classically Weyl-invariant action~\cite{BST},
\bena
S&=&\int d^{D} x \left\{\sqrt{-g}\bigl[\half \left((\partial \phi)^2-(\partial \vec{\chi})^2\right)\right.\label{eq1}
\\&&+ \left.\frac{(D-2)}{8(D-1)} (\phi^2 -\vec{\chi}^2)R - \rho\left(\frac{|J|}{\sqrt{-g}}\right)\bigr]-J^\mu \partial_\mu\tilde\varphi\right\}\,.\nonumber
\eena
The fluid is characterized by a densitized particle number flux $J^\mu$ with $|J|\equiv\sqrt{-g_{\mu\nu}J^\mu J^\nu}=\sqrt{-g}n$, where $n$ is the particle number density and $\rho(n)$ is the energy density~\cite{Brown}. The Lagrange multiplier $\tilde\varphi$ enforces particle number conservation: for a homogeneous, isentropic fluid there are no additional constraints.  The action (\ref{eq1}) is invariant under local Weyl transformations. While $\phi$  has the ``wrong sign'' kinetic term, there is no physical ghost: one can go to ``Einstein gauge'' where $\phi^2-\vec\chi^2$ is a constant, obtaining Einstein gravity coupled to scalars with positive kinetic energy. However, other Weyl gauges may be more convenient. 

{\em Quantum mechanics of a cosmological bounce.}---For homogeneous, isotropic cosmologies, one can choose a Weyl gauge in which the metric is static and the scalars $(\phi,\vec{\chi})$ encode all of the dynamics. While the metric is nonsingular in this gauge, the theory is still problematic because the effective Planck mass, given by the coefficient of $R$, can vanish, so that gravity becomes strongly coupled. Our strategy is to first identify this singularity in the quantum propagator, and then understand how to analytically continue around it. Our key assumption, which we shall test in various calculations, is that there are no singularities obstructing such a continuation. We set $D=4$ unless otherwise stated.

Fixing the metric to $ds^2=-N^2(t)dt^2+h_{ij}dx^idx^j$, where $h_{ij}$ is a metric of constant three-curvature $R^{(3)}=6\kappa$, Eq.~(\ref{eq1}) reduces to
\ben
S=V_0\int dt\,\bigl[\frac{\dot{\vec{\chi}}^2-\dot{\phi}^2}{2 N}+N\left(\frac{\kappa}{2}(\phi^2-\vec{\chi}^2) -\rho\right)-\tilde\varphi \dot{n}  \bigr]\,,
\label{eq2}
\een
where $V_0=\int d^3 x \,\sqrt{h}$ is the comoving spatial volume (which we take to be finite) and $n\propto \rho^{\frac{3}{4}}$. Define $x^{\alpha}:=(2\rho)^{-1/2}(\phi,\vec\chi)$, with $\alpha=0,\ldots,M$. Then, with $\eta_{\alpha \beta}={\rm diag} (-1,1,1\dots)$ and $m=2V_0\rho$, which is a coordinate scalar, Eq.~(\ref{eq2}) becomes 
\ben
S= \int dt \bigl[\frac{m}{2}\left(\frac{1}{N}\dot{x}^\alpha \dot{x}_\alpha-N(\kappa \,x^\alpha x_\alpha+1)\right)-\varphi\dot{m}\bigr]\,,
\label{eq3}
\een
with $\varphi:=\tilde\varphi V_0\,(dn/dm)$. Equation~(\ref{eq3}) is the action for a relativistic oscillator ($\kappa>0$), free particle ($\kappa=0$), or ``upside-down'' oscillator ($\kappa<0$), with  mass $m$, extending earlier work (see, e.g., Ref.~\cite{AT}) for the case $\rho=0$.

From Eq.~(\ref{eq1}), the effective Newton's constant is fixed by $(\phi^2-\vec\chi^2)= -2 \rho x^2$ so that, for positive radiation density, there are two timelike regions of superspace, with $x^2<0$ and $x^0>0$ or $x^0<0$, respectively, describing ``gravity," and a spacelike region, $x^2>0$, describing ``antigravity." The presence of radiation allows real solutions which pass smoothly from a ``gravity" region through an ``antigravity" region and back to a ``gravity" region, {\it i.e.}, {\it classical} bounces~\cite{BST}. Once anisotropies and inhomogeneities are included, generically there are no regular, real ``bounce" solutions; but there {\it are} regular, {\it complex} solutions which are deformations of the classical bounces. We claim these are legitimate saddle points of the path integral and provide a consistent semiclassical description of a {\it quantum} bounce. To describe these complex solutions, it is convenient to define the Einstein-frame scale factor $a$ by $x^\alpha=av^\alpha$ with $v$ an $(M+1)$-vector satisfying $v^2=-1$, so that real negative and positive values of $a$ represent the two ``gravity'' regions, while imaginary values of $a$ represent the ``antigravity'' region. We shall study generic complex solutions by analytically continuing $a$ into the complex plane and around the singularity at $a=0$. 

Let us begin with the propagator for homogeneous, isotropic universes. 

{\em Feynman propagator from path integral.} --- The propagator is given by a phase-space path integral \cite{hennteit},
\begin{widetext}
\ben
G(x,m| x',m') := \int \mathcal{D}x^\alpha \,\mathcal{D}P_\alpha\,\mathcal{D}m \,\mathcal{D}p_m\,\mathcal{D}N\, \exp\left(\im\int_{-1/2}^{1/2} dt\bigl[\dot{x}^\alpha P_\alpha+\dot{m}p_m-N\left(\frac{P_\alpha P^{\alpha}}{2 m}+ \frac{m}{2} \,  (\kappa \,x^\alpha x_\alpha+ 1)\right)\bigr]\right)\,.
\label{phsppathint}
\een
\end{widetext}
Equation~(\ref{phsppathint}) is obtained from a canonical analysis of Eq.~(\ref{eq2}) using Dirac's algorithm, after ``solving'' two second-class constraints to eliminate $\varphi$ and its momentum \cite{longpaper}. Fixing the gauge $\dot{N}=0$ allows us to replace the path integral over $N$ with an ordinary integral over proper time $\tau=N$ in this gauge \cite{halliwdw}. 
Integrating over $m$ and $p_m$ yields a delta function for $m$ conservation. The remaining path integrals are Gaussian and are computed exactly using the classical solution 
\ben
x(t)=\frac{x\sin\left[\sqrt{\kappa}\,\tau(t+\half)\right]+x' \sin\left[\sqrt{\kappa}\,\tau(\half-t)\right]}{\sin (\sqrt{\kappa}\, \tau)}\,.
\een
The final result is
\bena
G(x,m| x',m') &=& \im\delta(m-m')\int  d\tau\,\exp\left(-\im \frac{m}{2}\tau \right)\nonumber
\\&&\times\left(\frac{m \sqrt{\kappa}}{2\im\pi\sin(\sqrt{\kappa}\tau)}\right)^{(M+1)/2}
\label{fpropf}
\\&&\times\exp\left(\im m \sqrt{\kappa} \frac{(x^2+x'^2)\cos(\sqrt{\kappa}\tau)-2x\cdot x'}{2 \sin(\sqrt{\kappa}\tau)}\right)\,.\nonumber
\eena
For the Feynman propagator, $\tau$ runs from $0$ to $\infty$. We insert a convergence factor $-\epsilon \tau/(2m)$ in the exponent or,
alternatively, define the $\tau$-contour by steepest descent from the appropriate saddle point. The propagator has the usual short-distance singularities but is otherwise regular. It is defined by analytic continuation in $x^0$ (or $a$)  around these singularities through the half-plane in which it converges. For example, in extending the  amplitude from values for which 
$x^0-{x'}^{0}<|\vec{x}-\vec{x}'|$ to values for which $x^0-{x'}^{0}>|\vec{x}-\vec{x}'|$, we pass around the singularity in the lower-half $x^0$-plane. For $\kappa=0$, we obtain the massive free-particle propagator on flat spacetime, for which these analyticity properties are well known~\cite{longpaper,birdav}.

For homogenous, isotropic quantum cosmology with conformal matter, these calculations explicitly demonstrate that the Feynman propagator is perfectly regular at the ``big bang singularity" $x^2=0$ and that,  since the path integrals are Gaussian, the semiclassical approximation is exact. For a large universe, the background is ``heavy": there is little quantum spreading or backreaction from perturbations~\cite{rovelliwe}. More explicitly, provided radiation dominates, the action associated with its density [the first exponent in Eq.~(\ref{fpropf})] may be expressed as $ (3/8\pi G) H_E V_E$, where $H_E$ and $V_E$  are the Hubble constant and three-volume in Einstein gauge, and $G$ is Newton's constant. We shall perform our analytic continuations along complex paths for which this quantity increases at a rate sufficient to maintain the Wentzel--Kramers--Brillouin (WKB) expansion.  

We now turn to anisotropic, spatially flat cosmologies. Again, the Feynman propagator will be defined by analytic continuation around its singularities.

{\em Anisotropies.}---Consider a spatially flat, anisotropic metric in a conformal gauge where the determinant of the spatial metric is static,
\ben
ds^2=-N^2(t)dt^2+\sum_{i=1}^{D-1} e^{4\sqrt{\frac{D-1}{D-2}}\,\lambda_i(t)}dx_i^2\,,\quad \sum_{i=1}^{D-1} \lambda_i=0
\een 
(restoring the dimension $D$). The action (\ref{eq3}) becomes
\ben
S= \int dt \bigl[\frac{m}{2}\left(\frac{1}{N}\left(\dot{x}^2-x^2\sum_i\dot{\lambda}_i^2\right)-N\right)-\varphi\dot{m}\bigr]\,.
\label{aniso1}
\een
This is again the action of a massive free particle, now moving in a curved ``superspace'' metric $dx^2-x^2\,\sum_i d\lambda_i^2$, with a Lorentzian signature for $a^2=-x^2>0$. Recall, $x^\alpha=av^\alpha$, with $v^2=-1$, so the vector $v$ parameterizes a unit hyperboloid $H^M$. Taking into account the constraint $\sum_i  \lambda_i=0$, the $\lambda_i$ parameterize ${\mathbb{R}}^{D-2}$. Although the path integration is hard to perform directly, there is sufficient symmetry present for the {\it differential} equation that the propagator satisfies, the Wheeler--DeWitt equation, to determine it completely. Fourier transforming to the conserved momenta on $H^M\times{\mathbb{R}}^{D-2}$, only the dependence on $a$ remains to be determined. This, however, is fixed uniquely by the Wheeler--De Witt equation and analyticity properties. The Wheeler--DeWitt equation reads
\ben
\hat{\cal O}_aG(a,m|a',m')=-2\im m\,a^{-(M+D-2)}\delta(m-m')\delta(a-a')\,,
\label{wdw2}
\een
where 
\ben 
\hat{\cal O}_a\equiv \frac{d^2}{da^2}+\frac{M+D-2}{a}\frac{d}{da}-\frac{C}{a^2}+m^2,
\label{wdwop}
\een
with $C\equiv-\frac{1}{4}(M-1)^2-\zeta^2-k_A^2+\xi(D-2)(2M+D-3)$. Here, $\zeta$ is the momentum on $H^M$, $k_A$ that on the space of anisotropies, and $\xi$ is a parameter governing the ordering ambiguity identified by DeWitt~\cite{DeWitt} and clarified (in the phase-space path integral) by Kucha\v{r}~\cite{Kuchar}. Halliwell has made a strong case that $\xi$ should be taken to be the conformal coupling on superspace,  and further shown that this is consistent with the DeWitt inner product~\cite{halliwdw}. Formally, setting $D=2$ and $k_A=0$ in Eq.~(\ref{wdwop}) removes the anisotropy degrees of freedom and reduces the equation to that for the isotropic, flat case. 

The Feynman propagator is obtained by the Wronskian method from the positive and negative frequency modes, defined to be the solutions to $\hat{\cal O}_a\Psi(a)=0$ which are
regular in the lower- and upper-half complex $a$-plane, respectively. Up to normalization, they are given by 
\ben
 \Psi_{+,-}(a) =a^{-(M+D-3)/2}H_\nu^{(2,1)}(ma),
 \label{modes}
 \een
 where $\nu=\half\sqrt{4C+(M+D-3)^2}$, and $H_\nu^{(1,2)}$ are the Hankel functions of the first and second kind. The propagator is 
\bena
G(&a&,m|a',m')=\frac{\pi}{2} m\,\delta(m-m') (aa')^{-(M+D-3)/2} \cr
&&\times \left(H_\nu^{(1)}(ma')H_\nu^{(2)}(ma)\theta(a-a')+(a\leftrightarrow a')\right)\,.
\label{anisoprop}
\eena
For the flat, isotropic case, we obtain the free-particle propagator in the $(M+1)$-dimensional Minkowski space of fields $(\phi,\vec{\chi})$; for $M=0$, $G=\delta(m-m')e^{-\im m|a-a'|}$.

The modes appearing in the propagator have remarkable properties: an incoming positive-frequency mode continues to an outgoing positive-frequency mode, without even a phase shift. This surprising behavior follows from these facts: (i) the positive- (negative-)frequency modes are real and decay exponentially as $a$ runs to negative- (positive-)imaginary values and (ii) the WKB approximation holds at large $|a|$, throughout the lower-half (upper-half) complex $a$-plane. Thus, the modes continue to the positive or negative real $a$-axis where, by Schwarz reflection about the imaginary $a$-axis, their form is identical. The behavior of the DeWitt inner product is also of interest. In our example, where only the $a$ dependence is nontrivial,  this reduces to $\langle1|2\rangle\equiv a^{M+D-2}\Psi_1^* (a) \im \overleftrightarrow{\partial_a} \Psi_2(a)$.  If we normalize the positive- and negative-frequency modes at negative $a$, their norm is preserved as we pass to positive $a$, again by Schwarz reflection through Re$(a)=0$. 

It may seem strange that the singular potential in $\hat{\cal O}_a$ has no effect on scattering states. However, the classical theory echoes this behavior. Consider the Hamiltonian $H=N\half(p_a^2+C/a^2-m^2)$, with $C$ a constant and $N$ a Lagrange multiplier imposing $H=0$.  For $C>0$ (which can only occur when anisotropies are present), the classical solutions bounce and are nonsingular, 
\ben
a^2(\tau)=(\tau-\tau_0)^2+C/m^2,
\label{classaniso}
\een
with $\tau=N t$ and $\tau_0$ an arbitrary constant. We have $a(\tau)\approx (\tau-\tau_0)$ at large negative or positive $\tau$. The potential introduces no net time advance or delay: it slows the trajectory but also causes $a$ to bounce sooner.  For $C<0$,  the classical solutions are singular but become well defined if one gives the solutions an infinitesimal imaginary part. As  $a$ runs in from $-\infty$ and approaches the origin, it turns down the imaginary $a$-axis to $a=-\im \sqrt{-C}/m$ before heading back to the origin and out to $+\infty$. The speed-up due to the potential and the delay from the excursion into imaginary $a$ cancel. This excursion represents the ``antigravity" phase described in Ref.~\cite{BST}. This suggests that the ``antigravity'' regime is a consequence of trying to make the quantum behavior look classical, and that it should not be taken literally as a new classical phase.

Noting that, for $C>0$, one of the two modes diverges at $a=0$, some authors advocate setting $\Psi(0)=0$.  Effectively this renders $\Psi$ real, but (i) a real $\Psi$ cannot describe a state of nonzero $a$-momentum, i.e., an expanding or a contracting universe, (ii) the DeWitt norm is zero, so there is no meaningful notion of unitarity, and (iii) the correspondence principle is violated, due to a phase shift in the reflected mode which (as is easily seen) is independent of $m$. Furthermore, as $k_A$ is increased, $C$ goes negative. For small $C<0$, both solutions vanish at $a=0$, so there would be a jump in the allowed modes. At larger $C<0$ their prescription again gives a phase shift which violates the correspondence principle. None of these problems occur with our modes. 

{\em Inhomogeneities.}---Finally, we tackle inhomogeneities in a perturbative expansion around a flat ($\kappa=0$) background. The amplitude between in and out states is calculated in the saddle point (semiclassical) expansion around the appropriate complex classical solution. At lowest order, the initial state for the perturbations is a Gaussian wave functional for the incoming adiabatic vacuum. The corresponding classical perturbation is the positive frequency mode: any mode mixing with negative frequency modes across the bounce signals particle production.  Here, for simplicity, we set $M=0$ and study only planar perturbations, taken to nonlinear order.

We solve the equations of motion following from Eq.~(\ref{eq1}) in Einstein gauge. The background metric is given in Eq.~(\ref{eq0}): since $a(\eta)\propto\eta$, we can equivalently analytically continue in $a$ or in conformal time $\eta$. Assuming planar symmetry, the perturbed metric is 
\bena
ds^2&\propto&\eta^2\bigl[(-1+2\epsilon\phi) d\eta^2 + \left(1+2\epsilon(\psi+\gamma)\right)dx^2\label{comov}
\\&& + \bigl(1+\epsilon(2\psi+\half h^T)\bigr)dy^2+ \bigl(1 + \epsilon(2\psi - \half h^T)\bigr)dz^2\bigr]\,,\nonumber
\eena
where $\phi$, $\psi$, $\gamma$, and $h^T$ are functions of $\eta$ and $x$ only. For simplicity, we set the second tensor mode in $g_{yz}$ to zero (its dynamics are analogous to those of $h^T$).

We adopt a coordinate system in which the radiation is at rest everywhere (comoving gauge); the radiation density is $\rho(\eta,x)=\rho_0(\eta)[1+\epsilon\,\delta_r(\eta,x)]$. We expand the perturbations and Einstein equations in powers of $\epsilon$: $\phi(\eta,x)=\sum_{n\ge 1}\epsilon^{n-1}\phi^{(n)}(\eta,x)$, etc. At order $\epsilon^n$, Einstein's equation for the tensors $h^{T(n)}$ is
\ben
\frac{\partial^2 h^{T(n)}}{\partial \eta^2} + \frac{2}{\eta}\frac{\partial h^{T(n)}}{\partial \eta} - \frac{\partial^2 h^{T(n)}}{\partial x^2} = J^n(\eta,x)\,,
\label{einstein}
\een
where $J^n$ is nonlinear in the lower-order perturbations. There are four Einstein equations analogous to Eq.~(\ref{einstein}) for the scalar perturbations $\delta_r$, $\phi$, $\psi$, and $\gamma$. In Ref.~\cite{longpaper}, we derive the tensor and scalar Green's functions and solve Einstein's equations order by order in $\epsilon$. 

Consider the nonlinear extension of positive frequency solutions of the linearized equations, of wave number $k_0$,
\ben
\psi^{(1)}=A\cos(k_0 x)\frac{e^{-\im k_0 \eta/\sqrt{3}}}{k_0 \eta}\,,\; h^{T(1)}=B\cos(k_0x)\frac{e^{-\im k_0 \eta}}{k_0 \eta}\,,
\een
with $\phi^{(1)}$, $\gamma^{(1)}$, and $\delta_r^{(1)}$ determined in terms of $\psi^{(1)}$. 

\begin{figure}[htp]
\includegraphics[scale=0.45]{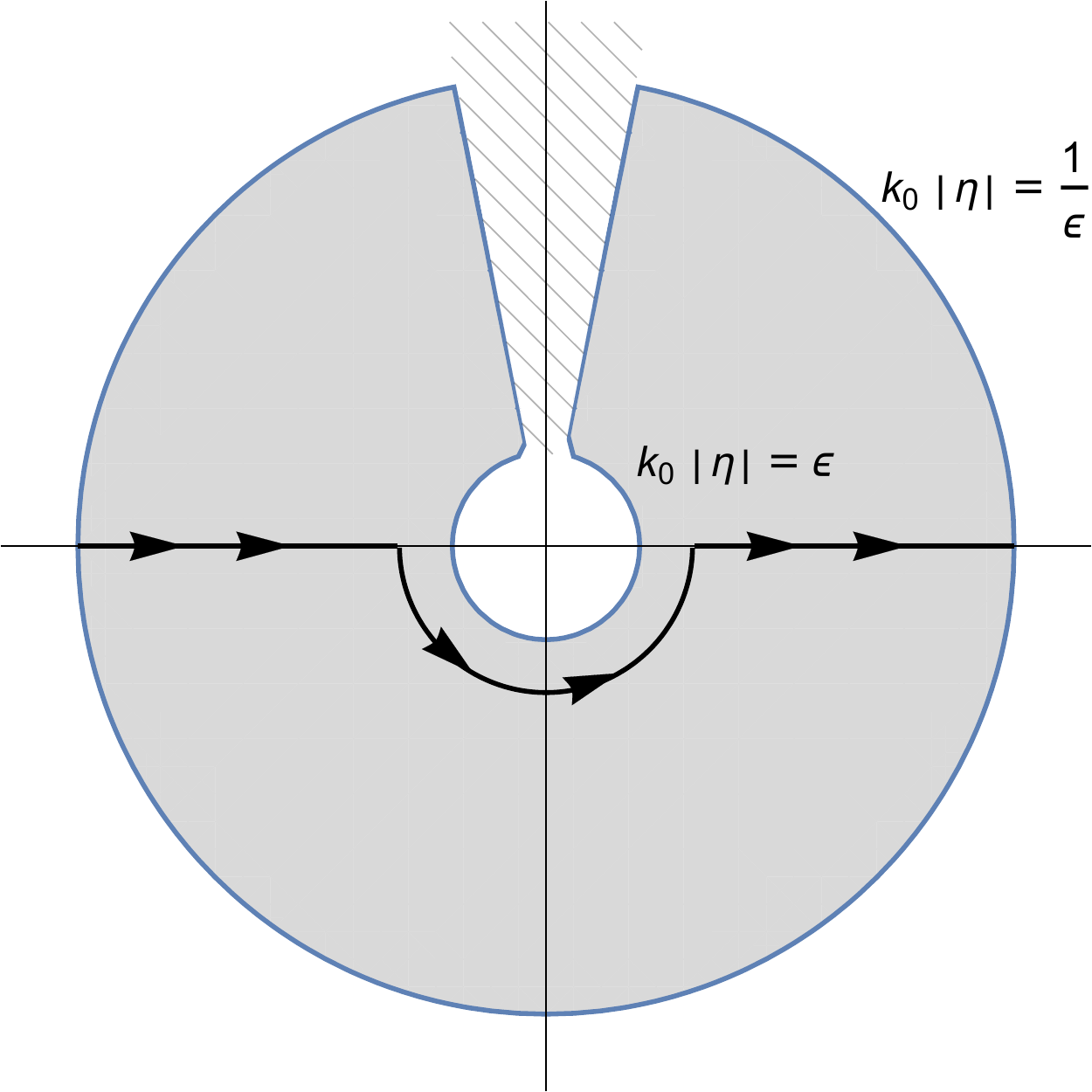}
\caption{Following a contour in the complex $\eta$-plane inside the annulus $\epsilon<k_0 |\eta|<1/\epsilon$. The dashed region indicates branch cuts in the upper half-plane.}
\label{contourfig}
\end{figure}

In Ref.~\cite{longpaper}, we give the positive frequency perturbations at second order, involving gamma functions and logarithms. After suitable definition of branch cuts, they are analytic in the lower-half $\eta$-plane, allowing us to avoid the singularity at $\eta=0$ (see Fig.~\ref{contourfig}), just as for the homogeneous background.  To calculate the quantum production of scalar or tensor perturbations across the bounce, we compute the leading terms of the nonlinear perturbations as $\eta\rightarrow\pm\infty$. At large $|\eta|$, $h^{T(2)}$ is
\ben
AB\,e^{-\im(1+\frac{1}{\sqrt{3}})k_0\eta}\frac{(27+16\sqrt{3})\cos(2k_0x)-6-5\sqrt{3}}{(21+11\sqrt{3})\im k_0\eta}+\ldots\,,\label{htasymp}
\een
where $\ldots$ are terms subleading in $(k_0\eta)$, and 
\ben
\psi^{(2)}(\eta,x) \sim \frac{A^2}{12} e^{-\frac{2}{\sqrt{3}}\im k_0 \eta}(1+2\cos(2k_0 x))+\ldots
\label{psiasymp}
\een
There is no mixing of positive and negative frequencies, and, hence, no particle production. The full functions $h^{T(2)}$ and $\psi^{(2)}$ decay exponentially for negative imaginary $\eta$. These properties extend to all orders in the perturbation expansion, unambiguously defining nonlinear positive frequency modes \cite{longpaper}.

On the real $\eta$-axis, $\psi^{(2)}$ does not decay at large $|\eta|$, indicating a breakdown of perturbation theory. Metric perturbations are gauge dependent; the ``gauge-invariant'' Newtonian potential $\Phi$ \cite{bardeen} is $\Phi^{(2)}\sim O(1/k_0\eta)$ at large $|\eta|$ at second order in $\epsilon$. Comparing with the first-order potential $\Phi^{(1)}\sim O(1/k_0^2\eta^2)$ still indicates a breakdown of the expansion when $k_0|\eta|\sim 1/\epsilon$. This phenomenon has a simple physical explanation. The radiation fluid is governed by nonlinear dynamics, eventually leading to shocks. A careful analysis \cite{fluidpaper} shows that perturbation theory breaks down when $k_0 |\eta| \sim 1/\epsilon$. Perturbation theory can thus be trusted for $\epsilon < k_0 |\eta| < 1/\epsilon$ (Fig.~\ref{contourfig}), where we obtain a perturbation expansion in $\epsilon$ for nonsingular, nonlinear solutions in the complex $\eta$-plane.

{\em Conclusions.} --- Our results indicate that a valid semiclassical approximation to quantum cosmology with conformal matter can be obtained from complex classical paths which avoid the classical big bang singularity. For homogeneous, isotropic backgrounds, and for anisotropic flat backgrounds, we computed the Feynman propagator exactly, showing its large $|a|$ behavior to be insensitive to the divergent potential introduced by anisotropies near the singularity. Finally, including inhomogeneities,  we found global, positive frequency modes; perturbation theory fails at {\it large} times, but this is physically well understood~\cite{fluidpaper}. Our investigations suggest the existence of a consistent and complete semiclassical description of a cosmological bounce, paving the way for detailed investigations of new, simpler and more predictive bouncing models.

The work of S.G. is supported by  the People Programme (Marie Curie Actions) of the European Union's Seventh Framework Programme (FP7/2007-2013) under REA grant agreement n$^{{\rm o}}$ 622339. Research at Perimeter Institute is supported by the Government of Canada through Industry Canada and by the Province of Ontario through the Ministry of Research and Innovation.


\begin{thebibliography}{99}

\bibitem{Friedmann} A.~Friedmann, ``\"Uber die Kr\"ummung des Raumes,'' {\em Z.\ Phys.} {\bf 10} (1922), 377--386.

\bibitem{LoopBounce} M.~Bojowald, ``Absence of singularity in loop quantum cosmology,'' {\em Phys.\ Rev.\ Lett.} {\bf 86} (2001), 5227--5230, \oarX{gr-qc/0102069}; A.~Ashtekar and P.~Singh, ``Loop Quantum Cosmology: A Status Report,'' {\em Class.\ Quant.\ Grav.} {\bf 28} (2011), 213001, \arX{1108.0893}.

\bibitem{Shomer} 
  A.~Shomer,
  ``A Pedagogical explanation for the non-renormalizability of gravity,''
  \arX{0709.3555}.

\bibitem{dewitt} B.~S.~DeWitt, ``Quantum Theory of Gravity. I. The Canonical Theory'', {\em Phys.\ Rev.} {\bf 160} (1967), 1113--1148.

\bibitem{BST}  I.~Bars, S.~H.~Chen, P.~J.~Steinhardt and N.~Turok,
  ``Antigravity and the big crunch/big bang transition,''
  {\em Phys.\ Lett.\ B} {\bf 715} (2012), 278--281, \arX{1112.2470};  I.~Bars, P.~J.~Steinhardt and N.~Turok, ``Local Conformal Symmetry in Physics and Cosmology,''
  {\em Phys.\ Rev.\ D} {\bf 89} (2014), 043515, \arX{1307.1848}.

\bibitem{Brown} D.~Brown, ``Action functionals for relativistic perfect fluids,'' {\em Class.\ Quant.\ Grav.} {\bf 10} (1993), 1579--1606, \oarX{gr-qc/9304026}.

\bibitem{AT} A.~Ashtekar and R.S.~Tate, ``An algebraic extension of Dirac quantization: Examples,'' {\em J.\ Math.\ Phys.} {\bf 35} (1994), 6434--6470, \oarX{gr-qc/9405073}.

\bibitem{hennteit} M.~Henneaux and C.~Teitelboim, {\em Quantization of Gauge Systems} (Princeton University Press, 1994).

\bibitem{longpaper} S.~Gielen and N.~Turok, ``Quantum cosmology with conformal matter,'' to appear.

\bibitem{halliwdw} J.~J.~Halliwell, ``Derivation of the Wheeler-DeWitt equation from a path integral for minisuperspace models,'' {\em Phys.\ Rev.\ D} {\bf 38} (1988), 2468--2481.

\bibitem{birdav} N.~D.~Birrell and P.~C.~W.~Davies, {\em Quantum Fields in Curved Space} (Cambridge University Press, 1982).

\bibitem{rovelliwe} C.~Rovelli and E.~Wilson-Ewing, ``Why are the effective equations of loop quantum cosmology so accurate?,'' {\em Phys.\ Rev.\ D} {\bf 90} (2014), 023538, \arX{1310.8654}.

\bibitem{DeWitt} B.~DeWitt, ``Dynamical Theory in Curved Spaces. I. A Review of the Classical and Quantum Action Principles,'' {\em Rev.\ Mod.\ Phys.} {\bf 29} (1957), 377--397, see p. 395. 

\bibitem{Kuchar} K.~Kucha\v{r}, ``Measure for measure: Covariant skeletonizations of phase space path integrals for systems moving on Riemannian manifolds'', {\em J.\ Math.\ Phys.} {\bf 24} (1983), 2122--2141.

\bibitem{bardeen} J.~M.~Bardeen, ``Gauge-invariant cosmological perturbations,'' {\em Phys.\ Rev.\ D} {\bf 22} (1980), 1882--1905.

\bibitem{fluidpaper} U.-L.~Pen and N.~Turok, ``Shocks in the Early Universe,'' \arX{1510.02985}.

\end{thebibliography}
\end{document}